\documentclass{aa501}
\usepackage{psfig}
\usepackage{lscape}
\usepackage{amsmath}

\newcommand{\MC}{\multicolumn}
\newcommand{\kms}{km~s$^{-1}$}
\newcommand{\funit}{10$^{-16}$~erg~cm$^{-2}$~s$^{-1}$}
\newcounter{qub}
\begin{document}

\title{6-m Telescope Spectroscopic Observations of
The Bubble Complex in NGC 6946}

\author {Yu.N.Efremov\inst{1} \and
S.A.Pustilnik\inst{2,3} \and
A.Y.Kniazev\inst{2,3,4} \and
B.G.Elmegreen\inst{5} \and
S.S.Larsen\inst{6} \and
E.J.Alfaro\inst{7} \and
P.W.Hodge\inst{8} \and
A.G.Pramsky\inst{2,3} \and
T.Richtler\inst{9}
}

\offprints{Yu. Efremov, \email{efremov@sai.msu.ru}}

\institute{
Sternberg Astronomical Institute of Moscow State University,
Universitetsky Prospect, 13, Moscow, 119899 Russia
\and  Special Astrophysical Observatory of RAS, Nizhnij Arkhyz,
Karachai-Circassia,  369167 Russia
\and Isaac Newton Institute of Chile, SAO Branch
\and Max Planck Institut f\"{u}r Astronomie, K\"{o}nigstuhl 17, D-69117,
Heidelberg, Germany
\and IBM Research Division, T.J. Watson Research Center,
    P.O. Box 218, Yorktown Heights, NY 10598, USA
\and UC Observatories / Lick Observatory, University of
  California, Santa Cruz, CA 95064, USA
\and Instituto de Astrofisica de Andalucia,  Granada, 18008 Spain
\and University of Washington, Seattle, USA
\and Grupo de Astronom\'{\i}a, Departamento de F\'{\i}sica,
     Casilla 160-C, Universidad de Concepci\'on, Concepci\'on, Chile
}

 \date{Received  September  9, 2001; accepted May 16, 2002}

 \abstract{
We describe the results of a long-slit spectroscopic study of an
unusual star complex in the nearby spiral galaxy NGC 6946 using
the SAO 6\,m telescope and the Keck 10\,m telescope. The complex
resembles a circular bubble 600 pc in diameter with a young super
star cluster (SSC) near the center. The kinematics of ionized gas
is studied through H$\alpha$ emission with several slit positions.
Position--velocity diagrams show two distinct features with high
speed motions. One is an irregularly shaped region to the east of
the SSC, $270$ pc in size, in which most of the H$\alpha$
emission is blue shifted by 120 km s$^{-1}$, and another is a 350
pc shell centered on the SSC with positive and negative velocity
shifts of 60 km s$^{-1}$.  Balmer and \ion{He}{i} absorption lines
in the SSC give an age of $12-13$ Myr, which is consistent with
the photometric age but significantly older than the kinematic
ages of the high speed regions.  The energetics of the SSC and its
interaction with the environment are considered.  The expansion
energies exceed $10^{52}$ ergs, but the power outputs from winds
and supernova in the SSC are large enough to account for this. The
intensities of Balmer, [\ion{N}{ii}], and [\ion{S}{ii}] emission
lines within and around the complex indicate that shock excitation
makes a significant contribution to the emission from the most
energetic region.}

\authorrunning{Y.Efremov et al.}

\titlerunning{Bubble Complex in NGC 6946}

\maketitle


 \section{Introduction}

Star complexes combine young clusters and associations along with slightly
older stars, like red supergiants and Cepheids, into vast groups 0.5
-- 1.0 kpc in size (Efremov \cite{Efremov95}).  
Most form by common processes related to spiral arm compression, gaseous
self-gravity, and interstellar turbulence, and represent
the largest scale in the hierarchy of structures related to star formation
(Elmegreen et al. \cite{Elmegreen00b}).
Others look very unusual with giant stellar arcs 
inside or at their borders making a near perfect segment of a circle.
These arc-like complexes are very rare (only a dozen are known), 
suggesting that some special process or event was involved with their
formation (Efremov \cite{Efremov01}, \cite{Efremov02}).

The first arc-shaped complex in a galaxy was noted by Westerlund \&
Mathewson (\cite{Westerlund66}) in the northeast part of the Large
Magellanic Cloud (LMC), inside the HI superbubble that is now known as
LMC4 from the survey by Meaburn (\cite{Meaburn80}). Hodge (\cite{Hodge67})
noted this arc also and found two others in the same region plus a similar
feature in the spiral galaxy NGC 6946.

The complex in NGC 6946 was recently rediscovered by
Larsen \& Richtler  (\cite{Larsen99}).
It has a massive young super star cluster (SSC) (Larsen \& Richtler
\cite{Larsen99}; Efremov \cite{Efremov99}) that is $\sim15$ Myr
old, and numerous other smaller clusters with about the same age
(Elmegreen et al. \cite{Elmegreen00a}).  The western edge is a 130 degree-long
arc with a radius of $\sim300$ pc.  There is no obvious pattern of
stellar age with radius inside the bubble.  Models of the color
magnitude diagram
observed with HST suggest that the SSC was triggered in a
centralized molecular cloud by compression from a surrounding OB
association; evolved remnants indicate that this association
formed 20 to 30 Myr ago and contained 1000 OB-type stars (Larsen et
al. \cite{Larsen02}).  Pieces of the cloud may be visible today as two $10^5$
M$_\odot$ dust arcs 150 pc to the north and northwest of the SSC
(Elmegreen et al. \cite{Elmegreen00a}). The HST photometry was also combined
with a velocity dispersion profile obtained at the Keck I telescope to
give an SSC mass of $\sim10^6$ M$_{\odot}$ and a central density
of $\sim10^4$ M$_{\odot}$~pc$^{-3}$ (Larsen et al. \cite{Larsen01}).

The origin and evolution of the complex with its peculiar circular
shape and massive cluster are unclear. More data on the
kinematics of the region are needed. For this reason, we observed
the complex with three spectroscopic slits using the 6\,m
telescope of the Special Astrophysical Observatory of the Russian
Academy of Sciences. We also studied another slit spectrum from  the
same Keck I observations as those used by Larsen et al. (\cite{Larsen01}).

Here we discuss the gas kinematics that
is evident from this spectroscopy, and we derive several cluster
ages from the Balmer lines.

\begin{table*}
\begin{center}
\caption{\label{Tab1} Journal of Observations}
\begin{tabular}{lllrcccccc} \\ \hline \hline
\MC{1}{c}{ Date }       &
\MC{1}{c}{ PA }         &
\MC{1}{c}{ Exposure }   &
\MC{1}{c}{ Wavelength } &
\MC{1}{c}{ Dispersion } &
\MC{1}{c}{ Slit }       &
\MC{1}{c}{ Seeing }     &
\MC{1}{c}{ Airmass } \\

    & &
\MC{1}{c}{ time [s] }    &
\MC{1}{c}{ Range [\AA] } &
\MC{1}{c}{ [\AA/pixel] } &
\MC{1}{c}{ [\arcsec] }    &
\MC{1}{c}{ [\arcsec] }    & \\

\MC{1}{c}{ (1) } &
\MC{1}{c}{ (2) } &
\MC{1}{c}{ (3) } &
\MC{1}{c}{ (4) } &
\MC{1}{c}{ (5) } &
\MC{1}{c}{ (6) } &
\MC{1}{c}{ (7) } &
\MC{1}{c}{ (8) } \\
\hline
\\[-0.3cm]
 26.07.2000 & -37.3  & 1x1800   & $6015-7250$  & 1.2 & 2.0 & 1.4 & 1.11 \\
 26.07.2000 & -37.3  & 3x1800   & $3850-5100$  & 1.2 & 2.0 & 1.4 & 1.06 \\
 26.07.2000 & ~83.3  & 2x1800   & $6015-7250$  & 1.2 & 2.0 & 1.4 & 1.11 \\
 26.07.2000 & ~29.5  & 1x1800   & $6015-7250$  & 1.2 & 2.0 & 1.4 & 1.16 \\
 28.07.2000 & ~83.3  & 3x1800   & $3850-5100$  & 1.2 & 2.0 & 2.2 & 1.06 \\
 28.07.2000 & ~29.5  & 3x1800   & $3850-5100$  & 1.2 & 2.0 & 2.4 & 1.05 \\
 28.07.2000 & ~29.5  & 2x1800   & $6015-7250$  & 1.2 & 2.0 & 2.7 & 1.12 \\
\hline \\[-0.2cm]
\end{tabular}
\end{center}
\end{table*}

 \section{Observations and data reduction}
\subsection{Long-slit spectroscopy with the 6\,m telescope}

Figure 1 shows two color images of the star complex.
The image on the left combines H$\alpha$, B, V, and R
from the Nordic Optical Telescope 
and the image on the right combines the same H$\alpha$
exposure with B, V, and R from HST. Four
slit positions used for this study are superposed. Tic marks on
the slits indicate the ranges of positions where fast H$\alpha$
emission is found (see below).

The observations for three of the slits were done on the 6\,m
telescope of the Special Astrophysical Observatory of the Russian
Academy of Sciences (SAO RAS) during 2 nights in July 2000 (see
Table~\ref{Tab1} for details). The Long-Slit spectrograph
(Afanasiev et al. \cite{Afanasiev95}) at the telescope prime focus
was equipped with a Photometrics CCD-detector with
1024$\times$1024 pixels and $24\times24\mu$m pixel size. The slit
length is 130\arcsec.  The spectra used a grating with 1302
grooves/mm and a dispersion of 1.2~\AA/pixel, giving a spectral
resolution of 3.2~\AA\ at H$\alpha$. The slit positions (see
Fig.~\ref{Fig:Slits}) were chosen to cross the SSC and various
characteristic features of this complex, including bright
\ion{H}{ii} regions and faint clusters near the SSC. The slit with
$PA=$ 83\degr\ points between the direction to the galaxy center
($PA=$ 70\degr) and the tangent direction to the local spiral arm
($PA \sim$93\degr). The slit position with $PA=$ 29\degr\ runs
across the arm.

The wavelength ranges for the spectra are given in Table~\ref{Tab1}.
A slit width of 2\arcsec\ was used in all cases.  The scale along the
slit is 0.39\arcsec~pixel$^{-1}$.  To better subtract the background
emission in the region of interest, we also got spectra of a blank
region $\sim$0.5\degr\ to the west of the target position. These 0.5-hour
spectra were in the same blue and red ranges as the target spectra.

Reference spectra of an Ar--Ne--He lamp were recorded before or after
each observation to provide  wavelength calibration.  Spectrophotometric
standard stars from Bohlin (\cite{Bohlin96}) were observed for
flux calibration. Observations have been conducted mainly under the
software package {\tt NICE} in MIDAS, described by Kniazev \& Shergin
(\cite{Kniazev95}).  To be confident that we observed the same parts of
the galaxy with the different set-ups on different nights, we employed
the differential method of pointing the telescope, which was described
in detail by Kniazev et al. (\cite{Kn_al_2001}).

Procedures of primary data reduction included cosmic-ray removal in
MIDAS,\footnote{MIDAS is an acronym for the European Southern Observatory
package --- Munich Image Data Analysis System.} and bias subtraction
and flat-field correction in IRAF. For the subsequent reduction of the
long-slit spectra, we used IRAF.\footnote{IRAF: the Image Reduction
and Analysis Facility is distributed by the National Optical Astronomy
Observatories, which is operated by the Association of Universities for
Research in Astronomy, In. (AURA) under cooperative agreement with the
National Science Foundation (NSF).}

After wavelength mapping and night sky subtraction using the blank
field frames, each 2D frame was corrected for atmospheric
extinction and flux calibrated. To derive the sensitivity curves,
we used the spectral energy distributions of standard stars.
Average sensitivity curves were produced for each observing night.
An example of the results of such a reduction, part of a
2D-spectrum, is shown in Fig.~\ref{Fig:2D-spectrum}. To obtain the
line-of-sight velocity distribution along the slit we used the
methods described in Zasov et al. (\cite{Zasov2000}).

\subsection{Keck-I telescope spectroscopy}

We supplemented our spectral analysis with high-resolution echelle
spectra from the Keck-I 10\,m telescope, obtained for an
independent project (Larsen et al. 2001) using the HIRES
spectograph (Vogt et al. \cite{Vogt94}). All details
of the observations  are described in that paper. For the present
analysis, we use only one of the echelle orders, containing
H$\alpha$ emission with a full extent of 14\arcsec\ along a slit
with a position angle of $-10^\circ$. Wavelength calibration was
performed at SAO, based on the two night-sky lines seen in this
order, namely H$\alpha$ and OH $\lambda$6568.779~\AA\ (e.g.,
Osterbrock et al. \cite{Osterbrock96}). This gives a scale of
0.048~\AA~pixel$^{-1}$, or 2.2~\kms~pixel$^{-1}$ at H$\alpha$.

\section{Observational results}

\subsection{The velocity distribution from the 6-m telescope data}

The spectra show H$\alpha$ emission over the whole extent of the
field, $\sim$110\arcsec--130\arcsec\ long (depending on the $PA$
of the slit), which is 3.2 to 3.8 kpc for a distance to NGC~6946
of 6.0 Mpc (Sharina et al.~\cite{Sharina97}).  Figure
\ref{Fig:Vel_distr_full} shows the H$\alpha$ intensity and
velocity measured with the $83^\circ$ slit, figure
\ref{Fig:Vel_distr_full1} shows the intensity and velocity from
the $-37^\circ$ and $29^\circ$ slits, and figure
\ref{Fig:Vel_distr} shows all three slits at higher spatial
resolution centered on the SSC.  The most detailed picture of the
H$\alpha$ velocity is seen in the spectrum with a slit orientation
of $PA=83$\degr, which was obtained on the 1st night of
observations with a seeing of 1.4\arcsec.  The $-37^\circ$ slit
had the same seeing, and the $29^\circ$ slit had poorer seeing,
2.7\arcsec (see Table 1).  All of these position-velocity diagrams
are arranged in the same way: the left sides correspond to the W,
SW or NW edges of the region, respectively (this is opposite to
the orientation of the sky field in Fig.~\ref{Fig:Slits}).  The
vertical lines are the position of the SSC.  The background radial
velocity of NGC~6946 is 125~\kms\ (Bonnarel et
al.~\cite{Bonnarel86,Bonnarel88}) in the region near the SSC. This
velocity is denoted by a dotted line in Figures
\ref{Fig:Vel_distr_full} and \ref{Fig:Vel_distr_full1}.

The H$\alpha$ velocity is disturbed by $\pm10$ \kms\ at several
places where there are enhancements in the H$\alpha$ emission. For
example, the velocity has positive and negative excursions at each
of the H$\alpha$ intensity peaks at positions 30\arcsec,
40\arcsec, and 50\arcsec\ in Fig.~\ref{Fig:Vel_distr_full}. These
disturbances could be unresolved shells caused by winds and old SN
explosions (for a review of the bubble/shell phenomenon, see
Tenorio-Tagle \& Bodenheimer \cite{TTB88}). At a distance of
40\arcsec\ in Figure ~\ref{Fig:Vel_distr_full}, both sides of what
appears to be an ionized shell in the top panel are also visible
in the middle panel at plus and minus velocities around the
average.  The negative velocity side of this shell, lying to the
right in the figure, is brighter than the positive velocity side.
Asymmetric H$\alpha$ shells were also found by
Martin~(\cite{Martin96, Martin98}, and see the discussion of the
``circumcluster'' shell below). At the spectral resolution of the
6\,m telescope, most small shells are unresolved and we see only
the weighted average velocity in each slit position.

There is another positive velocity excursion at the distance of
10\arcsec\ in the $29^\circ$ slit. This corresponds to a faint
filament seen in the H$\alpha$-image in Fig.~\ref{Fig:Slits}.

All three slits show a large, gradual, positive velocity
perturbation around the systemic speed of $125$ \kms for this
region.  It extends from about $-10$\arcsec\ to 20\arcsec\ in all
the slits, and has an amplitude of $40$ to $50$~\kms. This feature
is not a coherent shell since a receding part is not seen. It
could be a systematic velocity pattern from the galaxy rotation
curve and spiral arm flow, although the complex is at the end of a
short arm, not inside a main arm (see the whole galaxy image in
Elmegreen et al. \cite{Elmegreen00a}).

In the central part of the $83^\circ$ slit there is a
$-120$~\kms\
swing in the ionized gas velocity from 170 \kms\ to 50 \kms.
We
denote this feature as the ``fast'' expansion. The center of the
fast expansion is displaced to the east of the SSC by
$\sim$7\arcsec, or  $\sim$200 pc, and the extent of it is
$\sim$9\arcsec, or $260$ pc (the tic marks on the $83^\circ$ slit
in Figure \ref{Fig:Slits} indicate this extent). The center is
near the position of a cluster that has an age of $\sim$30 Myr,
labeled "33" in Fig. 4 of Elmegreen et al. (\cite{Elmegreen00a})
and $\#$502 in Fig. 1 of Larsen et al. (2002); this
cluster is seen in Fig. \ref{Fig:Slits} here, slightly above the
$83^\circ$ line, midway between the ticmarks). This cluster has an
absolute visual magnitude of M$_V=-8.65$.

The fast expansion is also present in the $PA=$29\degr\ slit
(Figure \ref{Fig:Vel_distr_full1}), where the amplitude of the
negative motion is slightly smaller, $-80$~\kms, and the full
extent of the region is larger, from 4\arcsec\ to 21 \arcsec\
along the slit. The transverse distance between the $83^\circ$ and
the $29^\circ$ slits at this position is comparable to the size of
the disturbance in each slit, so the center of the fast expansion
is somewhere between the two position angles, east of the SSC.

The H$\alpha$ intensities and velocities in the central parts of
the three slits are shown on an expanded scale in Figure~\ref{Fig:Vel_distr}.
The fast expansion down to a radial velocity
of 50~\kms\ is clearly seen in the $83^\circ$ slit at the top of
the figure, and partially seen in the $29^\circ$ slit in the
middle panel. Other agitated emission occurs at the position of
the SSC (the vertical line) in the $-37^\circ$ slit. This is the
only 6-m telescope slit that shows peculiar motion at the position
of the SSC; the $-10^\circ$ slit used for the Keck data shows it
better, as discussed in the next subsection.

The velocity-position profiles for the $83^\circ$ and $29^\circ$
slits in Figure ~\ref{Fig:Vel_distr} show a broad range of
velocities in the center of the fast emission.  The $-37^\circ$
slit shows the same broad range in the center of the emission near
the SSC.  These dispersions in the centroid velocities of the
emission lines correspond to broadenings in the actual line
widths, suggesting that there are either multiple components or
heightened turbulence.  The velocity resolution of the instrument
corresponds to a FWHM of $\sim145$ \kms\ (3.2 \AA), based on the
nearby reference spectrum and on night sky lines. Outside the
complex, the linewidths are indistinguishable from the
instrumental resolution. Within the complex, broadening over the
instrumental profile varies between 10\% and 20\%, and reaches a
maximum of $\sim25$\% near the SSC. These broadenings correspond to
Gaussian velocity dispersions of $\sim46$ \kms near the SSC, and
between 30 and 40 \kms\ in other parts of the complex out to
$10-15$\arcsec\ from the SSC, depending on direction. The high
dispersion near the SSC is consistent with the Keck telescope
data, which show an expanding shell there, as shown next.

\subsection{The velocity distribution from the Keck telescope data}

The 2D high resolution Keck-I echelle spectrum has greater sensitivity and
better seeing (0.9 \arcsec) than the other spectra, so it is useful for
faint small features.  We only observed near the SSC, however, so many
of the small velocity irregularities that occur in the 6-m telescope
slits cannot be seen with the Keck spectrum, nor can the large-scale
irregularities such as the fast expansion, which is comparable in size
to the Keck slit (14 \arcsec).

The Keck spectrum in 
Figure \ref{Fig:Vel_distr_Keck} shows another peculiar velocity
feature in the Keck data.  The velocity amplitude is $\sim$60
\kms\ and the diameter is $\sim12$\arcsec, comparable to the fast
feature seen in Fig.~\ref{Fig:Vel_distr_full}.  The position angle
of this slit is $-10^\circ$, which is similar to that of the
$-37^\circ$ slit taken with the 6-m telescope.

The emission in the Keck spectrum has both positive and negative
velocity components and extends on both sides of the SSC. It has
all the characteristics of a fragmented shell with a radius of
6\arcsec ($=170$ pc) expanding at $\sim60$ \kms.  In what follows,
we call it the ``circumcluster shell''. It is rather
inhomogeneous, with slightly brighter emission located in the NNW
sector (top of figure). In the SSE sector we see two separate
knots near the relative velocity $V=0$~\kms. The innermost knot
may be a dense ionized cloud inside the shell. The high-velocity
tail of lower intensity in the SSE sector of the position-velocity
diagram, stretching outward with relative velocities down to $-100$~\kms,
could be a piece of the fast expansion seen in the other slits.

The $\sim$6\arcsec\ radius of the circumcluster shell is comparable
to the distance between the SSC and the two dark clouds in the north
(cf. Fig. \ref{Fig:Slits}).  These clouds have an arched or shell-like
appearance centered on the SSC, suggesting that the H$\alpha$
circumcluster shell is the same as the partial dust shell.

\subsection{H$\beta$ emission}

The H$\beta$ emission inside the complex is in general
significantly fainter than the H$\alpha$.  Within the radii where
the fast expansion is seen in H$\alpha$, no H$\beta$ emission is
detected for all three positions of the slit.
This is not unreasonable considering the noise.
The Balmer
decrement from extinction also decreases H$\beta$
relative to H$\alpha$.
The extinction is at least partly caused by the Milky
Way's extinction at the position of NGC~6946 ($l^{II}=$95.7\degr,
$b^{II}=$11.7\degr), which is $A_B \sim$1\fm5 according to NED.
This implies a difference in the obscuration between H$\alpha$ and
H$\beta$ of $\sim$0\fm4, or a factor of $\sim$1.5.
The H$\alpha$/H$\beta$ ratio then changes from the standard
value of 2.88 for $T_e=10000$K without extinction to 4.3 with extinction.
The velocity field in H$\beta$
is more noisy in the W, NW and SW directions; the positive wave in
the velocity curve is well traced only in the E, NE, and SE sides.

\subsection{Physical conditions in the Complex}

Determination of the physical conditions inside the star formation complex
is difficult because of the intricate structure of ionized gas and the
many sources of excitation and shocks. Little emission is seen in the
[\ion{O}{iii}] lines $\lambda\lambda$4959,5007~\AA, and there is no
detectable emission in [\ion{O}{iii}] $\lambda$4363~\AA.

In Fig.~\ref{Fig:line_ratio} we show the profiles of the line
intensity ratios along the slit for $PA=$ 83\degr. The general
behaviour of these profiles can be summarized as follows. In the
close environment of the complex, the ratios are more or less
typical of high-metallicity, low-excitation \ion{H}{ii} regions
excited by massive stars.  Over a larger region and especially
inside the fast expansion, there is some increase of line ratios,
probably indicating a contribution from shock excitation. The
clear enhancement of line intensity ratios on both E and W
peripheries of the complex, in the regions of very low H$\alpha$
emission (with emission measure  $EM$ $\sim$15--30 pc cm$^{-6}$),
could be a Warm Ionized Medium (WIM) (see, e.g.,
Mathis~\cite{Mathis00}), a diffuse ionized gas (DIG), and/or a SNR
(e.g., Galarza et al. \cite{Galarza99}). Observations of WIM/DIG
in the Milky Way and M~31 display line intensity ratios similar to
those shown in the western, low-EM region (which are on the left
in Fig.~\ref{Fig:line_ratio}). However, the higher ratios seen in
the eastern, low-EM regions are more compatible with those
observed for SNRs (Galarza et al. \cite{Galarza99}).

\begin{table*}
\centering{ \caption{Equivalent widths of absorption lines for the
SSC and nearby clusters} \label{t:EW_CSC}
\begin{tabular}{llrr|llll} \hline
\rule{0pt}{10pt}
$\lambda_{0}$       &   SSC        &  Nearby  &  Nearby Blend       &Model
&Model         &Model         &Model          \\
      &              &clusters    & of 2 clusters      &6 Myr         &8 Myr
&12 Myr        &14 Myr         \\
      &              &  $PA=$ 83\degr\  &  $PA=-$37\degr\ &              &
&              &               \\
 \MC{1}{c}{(\AA)}  &   $EW$(\AA)    &  $EW$(\AA) &  $EW$(\AA)   &$EW$(\AA)
&$EW$(\AA)       &$EW$(\AA)       &$EW$(\AA)        \\
\MC{1}{c}{(1)}
&\MC{1}{c}{(2)}&\MC{1}{c}{(3)}&\MC{1}{c|}{(4)}&\MC{1}{c}{(5)}&\MC{1}{c}{(6)}
&\MC{1}{c}{(7)}&\MC{1}{c}{(8)} \\  \hline
4861\ H$\beta$\     & 3.0$\pm$0.3  &--1.9$\pm$0.3 &--3.9$\pm$0.6  & 4.6
& 4.0          & 4.6          & 5.8           \\
4340\ H$\gamma$\    & 5.8$\pm$0.6  & 4.7$\pm$0.4  & 3.9$\pm$0.8   & 3.8
& 4.3          & 4.5          & 5.6           \\
4101\ H$\delta$\    & 6.4$\pm$0.6  & 6.9$\pm$0.7  & 5.6$\pm$1.0   & 5.0
& 5.7          & 6.2          & 7.2           \\
3889\ H$_{8}$ \     & 3.6$\pm$0.6  & 2.9$\pm$0.7  & 4.3$\pm$1.2   & 3.8
& 4.6          & 4.8          & 5.6           \\ \hline
4922\ \ion{He}{i}\    & 0.8$\pm$0.2  & 0.8$\pm$0.2  &   ---         & 0.90
& 0.52         & 0.49         & 0.92          \\
4471\ \ion{He}{i}\    & 0.9$\pm$0.2  & 0.8$\pm$0.2  &   ---         & 0.87
& 0.80         & 0.94         & 1.1           \\
\hline
\MC{8}{l}{}\\
\MC{8}{l}{Model: Instantaneous SF at solar metallicity from Gonz\'{a}lez
Delgado et al. (1999)} \\
\MC{8}{l}{Minus in $EW$s of H$\beta$ means that this is the value for
emission.}
\end{tabular}
}
\end{table*}


\subsection{Spectra of the SSC}

The spectrum of the SSC is well exposed in both red and blue in a
total of 5.5 hours. Integrated over 7 pixels along the slit, it has a
signal-to-noise ratio (S/N) $\sim$60 at $\lambda$4800\AA\ and $\sim$30
near $\lambda$4000~\AA\ (see Fig.~\ref{Fig:cscspec}).  The Balmer
series in absorption from H$\beta$ to H$_{8}$ is well seen. Also several
\ion{He}{i} lines are seen in absorption at $\lambda\lambda$4388, 4471,
4921 and 5015~\AA. The radial velocity of the SSC determined from
these (and $\lambda$6678 of \ion{He}{i} in red) absorption lines is
150$\pm$10~\kms, which is close to the H$\alpha$ velocity of the ionized
gas in the position of the SSC.

The equivalent widths ($EW$s) of the Balmer absorption lines in
blue may be used to estimate the age of the SSC (Gonz\'{a}lez
Delgado et al. \cite{Rosa99}). The observed $EW$s are given in
Table~\ref{t:EW_CSC}. The measurements were made according to the
prescriptions given in Gonz\'{a}lez Delgado et al. in order  to minimize
systematic differences from their calibration. Typical $EW$ uncertainties
are $\sim$10\%.  Model results from Gonz\'{a}lez Delgado et al.  are given
on the right in the table. They assumed instantaneous star formation
with a Salpeter IMF having a mass range from $M_{low}$=1 $M_{\odot}$
to $M_{up}$=80 $M_{\odot}$.

Helium lines can also be used to obtain independent estimates of age,
but the uncertainties are higher by a factor of two because of their
lower $EW$s, and in the range of ages defined by the Balmer lines,
the $EW$s of He have a low sensitivity to age.

The SSC is associated with an emission-line region, so H$\beta$
absorption from the cluster can be contaminated by nebular emission. We
model the observed intensities of Balmer emission lines assuming Case
``B'' recombination at $T_e$=10000 K and a Milky Way extinction $A_B =$
1\fm5 (Schlegel et al.~\cite{Schlegel98}). With these data we derive
an $EW$ of H$\beta$ emission $\sim1.4\pm0.3$~\AA\ and a corrected value of
H$\beta$ absorption from the SSC of $EW$(H$\beta$) = 4.4$\pm$0.5~\AA.
According to Table ~\ref{t:EW_CSC}, this value of 4.4~\AA\ corresponds
to an age of either 6 Myr or 12 Myr.  The photometric age is $\sim15$
Myr (Larsen et al. \cite{Larsen01}).

For models with continuous star formation, the measured Balmer
EWs are less consistent with each other and correspond to ages of
$\sim$$30-50$~Myr. Such long time scales should be excluded
because then the $EW$ of  H$\beta$ emission is expected to be 30 times
higher than allowed by our data (Leitherer et al.
{\cite{Starburst99}).

\subsection{Spectra of other star clusters within the complex}

The continuum spectra of most other clusters are too faint to
measure EWs of their Balmer absorption lines. One of the best
spectra is of a faint object $\sim$5\arcsec\ to the west of the
SSC at $PA=$83\degr\ (seen toward negative positions in
in Fig.~\ref{Fig:2D-spectrum}).
This is the second object to the west of the SSC on the NOT
$I$-band image (marked 16 Myr in Fig. 4 of Elmegreen et
al.~(\cite{Elmegreen00a}). Because of poor seeing for this spectrum
($\sim2.5$\arcsec), it is highly probable that the
brighter object (marked ``19'' in that previous figure) 
$\sim$1.5\arcsec\ to the north of
object ``16'' gives a large contribution to the registered spectrum.
Its energy distribution is somewhat bluer  than that of SSC.

The Balmer lines H$\gamma$, H$\delta$ and H$_8$ of this nearby
cluster in the 83$^\circ$ slit
are well seen in absorption. Their EWs, uncorrected for
the underlying gas emission lines, are in
Table~\ref{t:EW_CSC} in the column denoted by the $83^\circ$
position angle. 
Besides the H$\beta$ line, which is heavily
contaminated by emission, the $EW$s of the other Balmer and
\ion{He}{i} lines are consistent within the uncertainties with
those for the SSC, which suggests that this fainter cluster has an
age that is about the same. However, its bluer spectral energy
distribution in UV implies that it may be a bit younger, with an
age of 6 to 9 Myr.

For the spectra of a fainter blend of two clusters in the $PA=-$37\degr\
slit, 
the contamination by underlying emission in H$\beta$ is even
larger. However, the $EW$s of absorption H$\delta$ and H$_8$ are
practically non-affected by nebular emission. The values in
Table~\ref{t:EW_CSC} are consistent with an age less than 14 Myr.

The $EW$ of H$\beta$ emission from the nebula can also be used as an
estimator of the nearby cluster age. 
The $EW$ of H$\beta$ from the vicinity of the
the two faint star clusters in the 83$^\circ$ slit
is 6  to 6.5~\AA. 
For an instantaneous starburst with Salpeter
IMF between $M_{low}$=1 $M_{\odot}$ and $M_{up}$=100 $M_{\odot}$ and
solar metallicity,
models in Leitherer et al. ({\cite{Starburst99}) 
give
an age of 7 Myr. For the faint pair of clusters seen in the spectrum
with $PA=-37$\degr, H$\beta$ emission is even stronger: $EW=8$ to
$8.5$~\AA, which corresponds to an age of 6.3 Myr.  These data
indicate that both nearby cluster groupings are younger than the bright
globular cluster. Younger isolated stars are also seen in the HST
data (Larsen et al. \cite{Larsen02}).

\subsection{Spectra of star clusters at the boundaries of the complex}
\label{border}

The same parameters can be used to estimate the ages of faint star
clusters that  delineate the arcs at the borders of the complex.
Their continuum is too faint to measure their Balmer absorption
lines, but well enough exposed to get an estimate of the $EW$s of
H$\beta$ nebular emission, and thus to get some insight about
their ages. As an example, we measured this parameter for the two
most visible faint clusters on both sides of the SSC in the blue
spectrum acquired with the slit at $PA=-$37\degr. In
Fig.~\ref{Fig:2D-spectrum_blue}, the positions of these clusters
are +12\arcsec\ (SE cluster) and $-12$\arcsec\ (NW cluster). The
$EW$s of their H$\beta$ emissions, calculated from 1D spectra
summed over 8 pixels along the slit (3.2\arcsec, where the cluster
continua are still visible) are, respectively, 22~\AA\ and 42~\AA.
After correction for the $EW$ of H$\beta$ absorption, which is
expected to be $\sim5$~\AA\ for an age between 5 and 13 Myr, we
derive true $EW$s of H$\beta$ nebular emission of 27~\AA\ and
47~\AA, respectively. From the models by Leitherer et al.
(\cite{Starburst99}) with the same IMF as above, we derive ages
for these star forming regions of 5.7 Myr for the SE cluster and
5.3 Myr for the NW cluster.  We have checked these small ages also
from independent data for $EW$(H$\alpha$), obtained from the red
spectrum. Their $EW$s  of 160 and 270~\AA\ give ages (Leitherer et
al. \cite{Starburst99}) consistent with those derived from the
$EW$(H$\beta$). As seen in Fig.~\ref{Fig:Vel_distr_full1} the full
extent of H$\alpha$ around these clusters is a bit larger than the
range over which the 1D spectrum was obtained. This leads to an
underestimate of related $EW$s of H$\beta$ and H$\alpha$ by 10
to 20\%, and to a decrease in the age estimates by $\sim5$ to
10\%.

Well-exposed spectra of these two clusters in red also show strong
lines of [\ion{N}{ii}]$\lambda\lambda$6548,6584 \AA. Their fluxes
relative to H$\alpha$ can be used to get some empirical estimate
of metallicities in the \ion{H}{ii} regions.  The relation:
\begin{equation}
\mathrm{12+\log (O/H) = 1.02 \cdot\log
(I([\ion{N}{ii}])/I(H\alpha))+9.36}
\end{equation}
was given by van Zee et al. (\cite{vanZee98}).
We get for these two clusters from our data 12+$\log (O/H)$=8.97 and 8.93,
respectively, with formal uncertainties of 0.2 dex. 
This oxygen abundance corresponds to the solar value (see,
e.g., Anders \& Grevesse \cite{Anders89}).

\section{Discussion}

We have undertaken a spectroscopic study of the velocity field of
ionized gas within and near a peculiar star complex in the SW part
of the giant spiral galaxy NGC~6946. Long-slit spectra taken at
three position angles separated by $\sim60^\circ$ sample the main
features of the velocity field. The spectra of the young central
star cluster and several fainter clusters were also obtained with
sufficiently high S/N ratio to measure the equivalent widths of
their Balmer absorption lines and/or the $EW$s of  H$\beta$
emission from the surrounding \ion{H}{ii} regions.  This allows us
to get information about their ages and motions relative to that
of the ionized gas.

The following parameters related to the processes taking place in
this stellar complex can be derived from the ionized gas
kinematics.

 \subsection{Characteristic timescales for the peculiar motions}

We estimate the characteristic timescales of the three large
velocity features connected with the complex. Two of them are seen
best on the position-velocity diagram in H$\alpha$ for the slit
position with $PA=$ 83\degr. The diameter of the slow perturbation
with a positive wave in the velocity curve and an amplitude of
$\sim50$~\kms\ (relative to the background velocity field) is
$\sim$25\arcsec\ or $\sim$730 pc. If we use the formula $v=0.6
R/t$ from Weaver et al.~(\cite{Weaver77}) for the speed of a shell
due to the pressure of a hot superbubble,  then the age of this
feature would be $0.6R/v=4.4$ Myr. For the Sedov (energy-conserving)
phase of a fast shell with an instantaneous energy injection, the
age would be $t=0.4 R/v=2.9$ Myr. For a spiral-arm flow, the
characteristic time scale would be $2R/v=15$ Myr.

The negative velocity feature 7\arcsec\ east of the SSC (the fast
expansion) has a smaller age. Its diameter in the
$83^\circ$ slit is 9\arcsec, or 260 pc,
and its current speed is 120 \kms\ (this is the amplitude of the
negative motion relative to the level of the slow perturbation
near the SSC). Applying the same formulae for winds and
supernovae, we get an age between 0.4 and 0.6 Myr. The
circumcluster shell has a diameter of $\sim12$\arcsec\ or $350$ pc,
and a speed of 60~\kms\, giving it an age of $\sim$1.5 Myr.

The fast expansion and circumcluster shell are younger than the SSC and its
neighbouring faint clusters.  Older regions with velocities of
$\leq$20~\kms\ would be more difficult to detect in such a
disturbed region. Many of the small velocity perturbations in the
spectra could be older.

\subsection{Kinetic energy of the perturbed regions}

The sizes and velocities of the expansion regions give their
kinetic energies, $E = 0.5 M v^{2}$ for mass $M$ and peculiar
velocity $v$. The mass of the fast expansion can be estimated from
the H$\alpha$ luminosity in the high velocity part of the emission
line. The H$\alpha$ surface brightness in the line is
$\sim$6$\times$10$^{-17}$ erg~cm$^{-2}$~s$^{-1}$ (square arcsec)$^{-1}$.
This corresponds to an emission measure of $<n_e^2>d =$ 30
pc$\cdot$cm$^{-6}$ for electron density $n_e$ and line of sight
depth $d$ in the shell (Marlowe et al. \cite{Marlowe95}; Osterbrock
\cite {Osterbrock89}).
The depth is unknown but is likely
to be less than 1/12 the radius $R=130$ pc of the expanding region,
or $d<11$ pc. This is because the ratio of the shocked density to
the ambient density is $R/3d$ for a spherical shell, and this
ratio exceeds 4 for a radiative shock. Then, $n_e>1.7$ cm$^{-3}$
and a spherical half-shell (we observe only the
approaching side) would have a mass of $2\pi R^2d n_e
\mu<6\times10^4$ M$_\odot$ for mean atomic weight including Helium
and heavy elements, $\mu=2.2\times10^{-24}$ gm. The corresponding
kinetic energy is $<9\times10^{51}$ ergs.

The initial energy of an explosion is greater than the mechanical
energy because of adiabatic and radiative losses during the
expansion. The initial energy of a supernova is
$E\sim5.3\times10^{43}n_0^{1.12}v^{1.40} R^{3.12}$ erg
for preshock density $n_0$ in
cm$^{-3}$, $v$ in km s$^{-1}$, and R in pc (Chevalier \cite{Chevalier74}). If
$n_0<0.4$ cm$^{-3}$, which is consistent with the shell density
limit and emission measure given above, then $E<6\times10^{52}$
ergs. If the expanding region is only a half-shell, then
$E<3\times10^{52}$ ergs.  Similarly, the injected energy for a
wind-driven shell in the snow-plow phase exceeds the mechanical
energy by a factor of $\sim5.1$ (Weaver et al. \cite{Weaver77}), making
$E<4.6\times10^{52}$ ergs.  The upper limits to these energies arise
from the uncertainty in shell thickness, given that the only
observation of mass is from an emission measure and this depends
on the square of the density. The actual energy is probably close
to these values because magnetic fields should dominate the pressure
in the swept-up gas, so the shell density will be close to the
minimum estimate. In any case, the energy limit is equivalent to
several tens of supernovae detonating within about a half million
years.

The energy of the circumcluster shell is $0.5Mv^2=2.4\times
10^{52}n_0$ erg for $v=60$ km s$^{-1}$ and $M = \left(4\pi/3\right)
 R^{3} n_0
m_{H}$ in a spherical shell with radius $R=170$ pc and pre-shock
density $n_0$. If $n_0\sim$0.1 atom cm$^{-1}$ as above, then the
kinetic energy is $2\times10^{51}$ ergs, and the injected energy is
$\sim5$ times larger for wind power.  These numbers imply a total
energy equivalent to $10$ SNe released in the last $\sim 1.5$ Myr.

The SSC has enough stars to power both the circumcluster shell and
the fast expansion. Figure \ref{Fig:energy} shows the wind and
supernova mechanical luminosities from a $10^6$ M$_\odot$ cluster like the
SSC over a period of 15 Myr, which is about the SSC age.  A Salpeter IMF
from 1 to 100 M$_\odot$ is assumed, using the data in
Starburst99 (Leitherer et al. {\cite{Starburst99},
scale down all luminosities by a factor of 2.5 for this
assumed cluster mass if the lower limit to the IMF is 0.1 M$_\odot$).  
The wind power (plotted as a dotted line) comes
mostly during the first 6 Myr, while the supernovae power (solid
line) starts after 3.5 Myr. Between 9.5 Myr and 12.5 Myr there is a
large burst of supernova power. This latter interval corresponds
to the expansion time of the circumcluster shell. The integrated
supernova energy during this time in the figure is $10^{55}$ ergs,
which is 5000 times the circumcluster shell energy. The SSC energy
from the last 0.5 Myr is $\sim2\times10^{53}$ ergs if we use the
low power part of Figure \ref{Fig:energy} that lies to the right
of the peak.  This is $\sim20$ times the mechanical energy
in the fast expansion, which has about this age. Evidently there
is more than enough power coming from the SSC to drive the
expansions observed spectroscopically.

A large part of the SSC
energy may have blown out of the galaxy.
The dust clouds in the north and northwest
and the high density gas behind the circular arc in the west
(Elmegreen et al. \cite{Elmegreen00a})
seem to have blocked the expansion in this direction, leaving
the winds and supernova energy relatively free
to move eastward.  This is where the lowest H$\alpha$
emission measure is, indicating a cleared cavity, and also
where the large negative velocities are found.
The lack of H$\alpha$ at the rest velocity suggests that
the whole galactic disk in this region is being pushed toward us at
$\sim100$ \kms.
That would imply either that the SSC is displaced slightly
toward the backside of the disk and has ionized and
accelerated the whole disk,
or that the backside part of
an ionized shell has already blown so far out of the disk,
away from the Lyman continuum source,
that its H$\alpha$ emission is now weak.

There is no reason to postulate any special energy source
for the high speed expansion other than supernovae
from the SSC. Some of the pre-supernova stars may have evaporated
from the SSC and moved to the east, aiding in the expansion there,
or a more exotic type of star or binary system could have
moved there, exploding as a hypernova (Paczynski \cite{Paczynski98};
Turatto et al. \cite{Turatto}).  The high energy and
short time scale of this expansion, $\sim10^{52}$ ergs in
$\sim0.5$ Myr, and the lack of any significant
clusters in the center of the fast region, suggests
that a lot of SSC energy got there quickly, as in the
ejected hypernova scenario. Hot gas could have moved
there quickly too. 

The possibility of a hypernova ejection from the SSC follows from recent
results for the dynamical evolution of star clusters.  Massive stars
sink to a cluster core rapidly and dynamical interactions lead to their
ejection. Kroupa (\cite{Kroupa00}) showed that during the first 10 to
50 Myr, between 10\% and 50\% of the most massive stars in a cluster
migrate to distances larger than the tidal diameter. This explains the
fact that most Cepheids (which are the most massive stars in a cluster)
occur in the halos (Efremov \cite{Efremov00}).  It also explains the
many massive stars found in the halo of the R136 cluster in the LMC
(Selman et al. \cite{Selman99}; Brandle et al. \cite{Brandle01}).

For a normal IMF, the most massive stars in the SSC should have a mass of
$\sim100$ $M_{\odot}$ (Larsen et al. \cite{Larsen01}), which is suitable
for a hypernova.  However, in the dense cores of clusters like this,
merging events should occur which lead to the formation of supermassive
and fast rotating stars or even black holes of a few hundred $M_{\odot}$
(Portegies Zwart et al. \cite{PZ99}; Matsushita et al. \cite{Matsushita};
Ebisuzaki et al. \cite{EMT01}).  These objects are the primary candidates for
hypernovae outbursts that may be connected with GRB events (Paczynski
\cite{Paczynski98}).  The most massive stars of a 12 Myr old cluster have
normally much lower mass. The suggestion therefore follows that the object
which might have formed the fast expansion in the hypernova scenario
resulted from a recent merger and ejection in the SSC.  
The possibility of the formation and ejection of GRB progenitors in dense
clusters was discussed in Efremov (\cite{Efremov00}).

\subsection{The relative velocities and positions of the perturbations and the
SSC}

The radial velocity of the SSC is $\sim$150~\kms. The kinematic center
of the slow perturbation is $\sim$125~\kms, and the kinematic center to
the sides of the fast expansion is $\sim$170 \kms.  The circumcluster
shell has the same kinematic center as the SSC ($\sim$150 \kms).
The SSC and the fast expansion have central velocities that are all
comparable to that of the surrounding gas. In Figure 5 of Bonnarel et al.
(\cite{Bonnarel88}), the region of the complex is within a large loop of
isolines between 125 and 140~\kms. The SSC and fast expansion velocities
are only slightly larger than the maximum average for the region, and the
slow perturbation is comparable to the minimum average.  The velocity
peculiarities of $\pm20$ km s$^{-1}$ are not strong enough to suggest
any particular scenario for the origin of the whole region. It lies near
the end of an arm segment or spur, where density wave flow speeds are
expected to be comparable to these velocity fluctuations.

The morphology and intensity of star formation in this region are unusual
so there might be an unusual explanation for its origin.  There is a
comet-like shape to the whole region, with a sharp circular edge in the
west and a more open, tail-like structure in the east.  
Figure \ref{Fig:bubble} shows an enhanced version from the NOT image. 
This shape could
have been triggered by the oblique impact of an extragalactic cloud,
from east to west.  The impact had to be oblique so that
the resulting disk velocities were not too different from the local rest
speed.  Cloud-disk collisions have been suggested for NGC 6946, but HVCs
and HI holes are only known in other regions (Boulanger \& Viallefond
\cite{Boulanger92}; Kamphuis \cite{Kamphuis93}).  There is no prominent
HI hole in the complex, only a small hole in the region of the fast
expansion east of the SSC, as seen in Figure 1f of Frick et al.
(\cite{Frick00}).
An oblique collision of a high-velocity cloud
with a magnetized gaseous disk results in MHD waves, gas vortices, and
Parker instabilities, but no prominent hole (Santill\'an et al.
\cite{Santi99}).
All of this, along with the resulting shock wave collisions, could
explain some the peculiarities of the complex, including the discrete
epochs of star formation, the dust lane across it, the circular Western
rim, and the small perturbations in the HII velocities.  The SSC could
then be the result of the shock wave interaction.  These possibilities
are considered in more detail elsewhere  (Efremov \cite{Efremov02}).

A problem with the impact theory is that the comet also looks like a
normal spiral arm spur with a continuum of blue stars inside, and there
are other similar spurs in this galaxy.  This makes the origin of the
complex look more like a spiral arm process than a random cloud impact.
Also, some of the asymmetric morphology could result from a high-pressure
expansion toward the east using energy from the SSC and a density gradient
toward the west (see the blister models in Icke \cite{Icke81}).  There is in
fact a higher density of HI to the west in the Frick et al. map, and there
is more extinction there in our optical images as well (see Elmegreen
et al. \cite{Elmegreen00a}).

\section{Conclusions}

Long-slit spectroscopy of a unique stellar complex in NGC~6946 has
shown several interesting features that shed light on this unusual
object. We draw the following conclusions:

\begin{enumerate}
\item{
Position-velocity diagrams of
H$\alpha$ emission in the complex indicate the
presence of three large-scale motions: a slow perturbation with a
full size of $700-800$ pc and a velocity of $\sim$50~\kms, a fast
expansion, with a size of 260 pc and a velocity of 120 \kms, and a
circumcluster shell with a size of 350 pc and a velocity of
60~\kms. The latter is centered in both position and velocity on
the super star cluster.  The fast expansion may be centered 200 pc
away from this cluster but still receive energy from it.
The large energy and short time scale for the fast
expansion suggests that an ejected hypernova might have been involved.
}

\item{
Well sampled spectra of the young globular cluster  show Balmer
and \ion{He}{i} absorption lines with radial velocity of $\sim$150
\kms, close to that of the ionized gas in this region. The EWs of
absorption lines and of the H$\beta$ emission line are used to
date the ages of the SSC and two nearby clusters. The age of the
SSC is 12$\pm$2 Myr, and those of the nearby fainter clusters are
significantly smaller, $\sim$7 Myr. Some clusters on the periphery
of the complex are even younger, with ages of $\sim$5.5 Myr. These
ages are in good agreement with those found from the UBV
photometry obtained with the HST (Larsen et al. \cite{Larsen02}).
The HST data also obtained ages for other clusters and found no
pattern with position. }

\item{The ratios of emission line intensities
([\ion{N}{ii}]6584/H$\alpha$ and
([\ion{S}{ii}]6716+6731)/H$\alpha$) indicate  that shock
excitation can contribute in the region of the fast expansion. }

\item{The overall  velocities  of the SSC and most of the \ion{H}{ii}
gas have no strong deviations from the rotation curve of NGC
6946.}
\end{enumerate}

\begin{acknowledgements}

This research has made use of the NASA/IPAC Extragalactic Database
which is operated by the Jet Propulsion Laboratory, California
Institute of Technology, under contract with the NASA. The use of Digitized
Sky Survey and NASA Astrophysics Data System  is gratefully
acknowledged. Yu.E. appreciates the  support from the RFBR,
grant 00-02-17804 and  from the
Conseil for the Support of the Scientific Schools, grant
00-15-96627.  B.G.E and S.L. were supported by HST grants
GO-08715.02-A and GO-08715.05-A.  EJA acknowledges partial support
from DGICYT trough grants PB97-1438-C02-02 and by Research and
Education Council of the Autonomous Government of Andalucia
(Spain). Thanks are due to the referee M.Rieke, who made several
useful suggestions which helped to improve the presentation of
results and discussion.
\end{acknowledgements}

\clearpage

\begin{figure}
{\centering
}
 \caption{The positions of the long slits from the 6-m telescope and the
 shorter slit from Keck I are superposed on a
combined image using H$\alpha$, B, V, and R from the Nordic
Optical Telescope (left) and a combined image with the same
H$\alpha$ from NOT and B, V, and R from HST (right). For the left-hand
image, North is up, East is to the left, and the full extent
East-West is 43 \arcsec.  For for the right-hand
image, North is indicated by the arrow and the full
extend left-right is 20.9 \arcsec. 
Tic marks
indicate ranges of positions where the centroid velocity of the
H$\alpha$ line shifts to large negative values.}
\label{Fig:Slits}
\end{figure}

\begin{figure}
\vspace{5.0cm}
{\centering
}
\vspace{-5.2cm}
\caption{The "blue" 2D-spectrum for $PA=-$37\degr\ corrected for
CCD sensitivity. The frame is an average of three frames with an
exposure time of 0.5 hour for each. The spectrum contains
H$\gamma$ $\lambda$4340 \AA, H$\beta$, and [\ion{O}{iii}]
$\lambda$5007 \AA\ emission lines, together with Balmer
absorptions. The SSC is at the ordinate $Y$=0\arcsec. The NW
direction corresponds to the negative ordinate. Two weaker, almost
overlapping clusters inside the complex are seen at the ordinate
$Y$=--5\arcsec\ to --10\arcsec. They are clear in the H$\alpha$
profile in Fig.~\ref{Fig:Vel_distr_full1}. There is weak blue
emission inside the complex in contrast to the outside region at
the ordinates $Y<$--15\arcsec\ and $Y>$15\arcsec.}
\label{Fig:2D-spectrum_blue}
\end{figure}

\begin{figure}
\vspace{-0.5cm}
{\centering \hspace*{3.0cm}
}
\caption{Part of the ``red'' 2D-spectrum for $PA=$ 83\degr\, which
contains H$\alpha$ (6562 \AA), [\ion{N}{ii}] $\lambda$6548,6584
\AA\, and [\ion{S}{ii}] $\lambda$6716,6730 \AA\ emission lines.
The central star cluster is at the ordinate $Y=$0\arcsec. The
direction to the West corresponds to negative ordinates. One more
rather bright cluster inside the complex is at the ordinate
$Y=$--5\arcsec. The arc-like form of H$\alpha$ is seen for the
ordinates between 0 and +10\arcsec (enlargement on right). This
corresponds to the high-velocity expansion with a negative displacement in
Fig~\ref{Fig:Vel_distr_full}. The same feature can be traced for
the lines [\ion{N}{ii}] $\lambda$6584 \AA\ and [\ion{S}{ii}]
$\lambda$6716 \AA. These frames were made from an average of two
frames with exposure times of 0.5 hour; each separate frame shows
the arc-like feature of H$\alpha$ near the ordinate $Y=$5\arcsec.
} \label{Fig:2D-spectrum}
\end{figure}

\begin{figure}[hbtp]
{\centering
} \caption{H$\alpha$ intensity and radial velocity distributions
along the slit with $PA=$ 83\degr\ in a region 130\arcsec\ (3.8
kpc) in extent. The vertical line at $X=$0 corresponds to the
position of the SSC, which is 4\arcsec\ SSE of the center of the
complex. The leftward direction in this plot corresponds to a
rightward direction in Figure 1 and in the sky. {\it Upper panel:}
The lines at the bottom and middle show on a reduced scale (1/26
of the scale shown on $Y$-axis) the general behavior of the
continuum intensity near H$\alpha$ and the flux of the
H$\alpha$-line, respectively. The upper line shows the profile of
H$\alpha$ flux along the slit with the continuum subtracted in
real counts. {\it Middle panel:} The velocity profile of the
H$\alpha$ line, corrected using the night-sky line [\ion{O}{i}]
$\lambda$6363 \AA. The horizontal line shows the background
velocity of NGC~6946 near the complex as outlined by H$\alpha$
data from Bonnarel et al.~(\cite{Bonnarel86,Bonnarel88}).
{\it Bottom panel:} The deviations
from the zero-position after wavelength transformation of the
night-sky line [\ion{O}{i}] $\lambda$6363 \AA, with a fourth-order
polynomial fit superimposed. This fitting curve was used to
correct the first pass velocity curve and the result of the
correction is shown in the middle panel (b). The r.m.s. of the
distribution relative to the fitted curve is $\sim$3 \kms. }
\label{Fig:Vel_distr_full}
\end{figure}

\begin{figure}[hbtp]
{\centering \hspace*{3.cm}
} \caption{The H$\alpha$ intensity and radial velocity
distributions along the slits with $PA=-$37\degr\ and $PA=$
29\degr, extending for 115\arcsec\ (3.3 kpc). All panels are
similar to those in Fig. \ref{Fig:Vel_distr_full}.}
\label{Fig:Vel_distr_full1}
\end{figure}

\begin{figure}[hbtp]
 {\centering
}
\caption{ {\it Upper panel of each figure:} The brightness profile of
the H$\alpha$ line + continuum (upper curve) and the brightness profile
of the continuum near the H$\alpha$ position (lower curve). {\it
Lower panel of each figure:} The velocity profile of the H$\alpha$
line.
The left hand parts of these
three plots correspond respectively to W, SW and NW directions (see for
reference Fig.~\ref{Fig:Slits}). For $PA=$ 29\degr, data
are shown for both the first (sharper curves near position of SSC) and second
nights. The significant effect of poor seeing is seen for the second
night. }
\label{Fig:Vel_distr}
\end{figure}

\begin{figure}[hbtp]
{\centering
} \caption{ A fragment of the Keck-I 2D echelle spectrum of the
SSC and its vicinity near H$\alpha$ with the 14\arcsec\ long slit at
$PA=-$10\degr. The SSC is at the position along the slit corresponding
to the ordinate  $Y$=0\arcsec. Positive values of ordinate correspond to
the NNW direction. The $X$-axis corresponds to the radial velocity of
H$\alpha$ emission relative to the SSC velocity, taken to be 150~\kms
from the SSC spectrum. A shell with a radius of $R\sim$6\arcsec\
and a velocity of 60~\kms\ is centered on the SSC
($V_\mathrm{kin.center}=$ 144~\kms). A lower intensity feature with
a radial velocity down to $-100$~\kms\ relative to the SSC stretches to
SSE. This is probably the fast expansion, seen in the middle
panel of Fig.~\ref{Fig:Vel_distr_full}. }
\label{Fig:Vel_distr_Keck}
\end{figure}

\begin{figure}
{\centering
}
\caption{ Line intensity ratios along the slit for $PA=$83\degr. {\it Top
to bottom:} a). Profile of H$\alpha$ flux in units \funit. Lower curve
is scaled by a factor of 1/10 to show bright regions. b). Profile of
[\ion{N}{ii}6584]/H$\alpha$ ratio. While the background level of this
parameter near the complex is $\sim$0.3, it increases to $\sim$0.4
in the region of slow perturbation, and rises up-to $\sim$0.5  inside
the fast expansion and up-to $0.6-0.8$ in the peripheral regions to
the E and W with the low level of H$\alpha$ emission.  c). Profile
of [\ion{S}{ii}]6716+6730/H$\alpha$ ratio. Its background level of
$\sim$0.3 shows little change within the slow perturbation, however large
variations are seen inside the fast expansion. The zero level of this
ratio is shown for non-detected [\ion{S}{ii}] lines. This ratio is also
enhanced up-to factors of $0.6-1.2$ in peripheral regions with faint
diffuse H$\alpha$ emission.  d). Profile of the density sensitive
[\ion{S}{ii}]6716/[\ion{S}{ii}]6730 R$_{[\ion{S}{ii}]}$ ratio.
For this ratio $\geq$ 1.5 (dot-dash line) n$_e <$ 10~cm$^{-3}$.  For
R$_{[\ion{S}{ii}]} =$1.4, 1.3, 1.2, 1.1, 1.0 and 0.9 respectively, n$_e
=$ 40, 130, 240, 380, 580 and 850~cm$^{-3}$. The typical uncertainty in
this ratio varies between 0.1 for the brightest features, as illustrated
in Fig.~\ref{Fig:2D-spectrum}, and $0.4-0.5$ for the lowest brightness
regions.  Inside the fast expansion, these lines are almost everywhere
undetectable, and the ratio is set to zero conditionally.  }
\label{Fig:line_ratio}
\end{figure}

\begin{figure}
{\centering \hspace*{1.5cm}
} \caption{Line intensity ratios along the slit for $PA=-$37\degr.
a) Distribution along the slit of the line flux (in units \funit)
for H$\alpha$.  The distribution of H$\beta$ is shown by the middle line,
and that of [\ion{O}{iii}] $\lambda$5007 \AA\ by the bottom line.
Other panels show line intensity ratios as in the previous figure.
The profile of the [\ion{O}{iii}] $\lambda$5007/H$\beta$ ratio is
shown in panel e).  }
\label{Fig:line_ratio_37}
\end{figure}

\begin{figure}
{\centering
}
\caption{ Line intensity ratios along the slit for $PA=$29\degr.
All panels are as in Figure~\ref{Fig:line_ratio}.  }
\label{Fig:line_ratio_29}
\end{figure}

\begin{figure}
{\centering
} \caption{ 1D composite spectrum of the SSC, obtained by
averaging of 1D
spectra for the slits with $PA=$29\degr\ and $PA=$83\degr.  
The equivalent widths of the Balmer and \ion{He}{i}
absorption lines are consistent with an age of the SSC equal to
$\sim$12 Myr (in the approximation of instantaneous star
formation). The signal-to-noise ratio of the composite spectrum as
a function of wavelength is shown in the bottom panel. }
\label{Fig:cscspec} \end{figure}

\begin{figure}
{\centering
} \caption{ \underline{\it Left panel:} Blue spectra of the SSC
(top) and nearby star cluster (bottom) at $PA=$ 83\degr\ . The
latter shows a steeper rise of its spectrum toward the UV in
comparison to that of SSC. The Balmer series is well seen in
absorption, as well as the \ion{He}{i} $\lambda$4471,4921 \AA\
lines. \underline{\it Right panel:} Red spectra of the SSC with
$PA=-$37\degr\ on the top and $PA=$ 83\degr\
in the middle; the spectrum of the nearby star cluster is
on the bottom. Emission from the surrounding ionized gas is
superposed on the absorption spectra of the star clusters.
\ion{He}{i} $\lambda$6678~\AA\ absorption in the spectrum of SSC
is seen. Redward of [\ion{O}{i}] $\lambda$6300~\AA\  and
$\lambda$6364~\AA\ non-perfect sky subtraction creates artificial
emission. The red spectrum of the SSC observed with $PA=-$37\degr\
is shifted up by +3.5 units for convenience. The difference
between the red spectrum of the SSC at $PA=-$37\degr\ and that at
$PA=$83\degr\ is less than 5\% in the range from $6000$ to $6800$
\AA\, but it jumps up to $\sim$20\% in the small region around
H$\alpha$, which is seen in absorption. However the fluxes of
emission H$\alpha$ line in these spectra differ by only $\sim$6\%.
} \label{Fig:clus_spec}
\end{figure}


\begin{figure}
{\centering
}
\caption{
The history of the mechanical luminosity of a $10^6$ M$_\odot$ cluster
in units of $10^{40}$ erg s$^{-1}$.  The dotted line is the
wind luminosity and the solid line is for supernova.
}
\label{Fig:energy}
\end{figure}

\begin{figure}
{\centering
}
\caption{
The Bubble complex from an enhanced version of the NOT 
color image.  Note the perfect semiring for the
Western rim.  The center of the drawn circle 
is noted by the small grey circle.}
\label{Fig:bubble}
\end{figure}


\begin{thebibliography}{99}
\bibitem[1995]{Afanasiev95}
 Afanasiev, V.L., Burenkov, A.N., Vlasyuk, V.V., Drabek, S.V. 1995, SAO
  RAS,  internal report, No.~234
\bibitem[1989]{Anders89} Anders, E., \& Grevesse, N. 1989,
    Geochim.Cosmochim.Acta, 53, 197
\bibitem[1996]{Bohlin96} Bohlin, R.C.  1996, AJ, 111, 1743
\bibitem[1986]{Bonnarel86} Bonnarel, F., Boulesteix, J., \& Marcelin, M.
  1986, A\&AS, 66, 149
\bibitem[1988]{Bonnarel88} Bonnarel, F., Boulesteix, J., Georgelin, Y.P.,
  et al. 1988, A\&A, 189, 59
\bibitem[1992]{Boulanger92} Boulanger, F., \& Viallefond, F. 1992, A\&A,
    266, 37
\bibitem[2001]{Brandle01} Brandle, B., Chernoff, D.F. \& Moffat A.F.J., 2001,
  in: E.Grebel, D.Geisler and
   D.Minniti, eds,. "Extragalactic star clusters", Proceed. IAU Symp. 207,
   in press.
\bibitem[1974]{Chevalier74} Chevalier, R. 1974, ApJ, 188, 501
\bibitem[2001]{EMT01} Ebisuzaki, T., Makino, J., Tsuru, T.G.,
   et al. 2001, ApJ, 562, 19
\bibitem[1995]{Efremov95} Efremov, Y.N. 1995, AJ, 110, 2757
\bibitem[1999]{Efremov99} Efremov, Y.N. 1999, Astron. Lett., 25, 100
\bibitem[2000]{Efremov00} Efremov, Y.N. 2000, Astron. Lett., 26, 558
\bibitem[2001]{Efremov01} Efremov, Y.N. 2001, Astron. Rep., 45, 769
\bibitem[2002]{Efremov02} Efremov, Y.N. 2002, Astron. Rep., in press
\bibitem[2000a]{Elmegreen00a} Elmegreen, B., Efremov, Y.N. \& Larsen, S.
   2000a,  ApJ, 535, 748
\bibitem[2000b]{Elmegreen00b}Elmegreen, B.G., Efremov, Y.N, Pudritz, R., \&
   Zinnecker, H. 2000b, In: Protostars and Protoplanets IV, eds. V.Mannings,
   A.P.Boss, and S.S.Russel,  Univ. Arizona Press, p. 179
\bibitem[2000]{Frick00} Frick, P., Beck, R., Shukurov, A., et al.
   2000, MNRAS, 318, 925
\bibitem[1999]{Galarza99} Galarza, V.C., Walterbos, R.A.M., \& Braun, R.
  1999, AJ, 118, 2775
\bibitem[1999]{Rosa99} Gonz\'{a}lez Delgado, R.M., Leitherer, C. \&
  Heckman, T.M.   1999, ApJS, 125, 489
\bibitem[1967]{Hodge67} Hodge, P. 1967, PASP, 79, 29
\bibitem[1981]{Icke81} Icke, V. 1981, ApJS, 45, 585
\bibitem[1993]{Kamphuis93} Kamphuis J.J., 1993, Ph.D. Thesis, Univ.
  of Groningen
\bibitem[1995]{Kniazev95} Kniazev, A.Y., \& Shergin, V.S. 1995, SAO RAS
  internal report No.~249, 1
\bibitem[2001]{Kn_al_2001} Kniazev, A.Y., Pustilnik, S.A., Pramsky, A.G.,
    \&  Ugryumov, A.V. 2001, A\&A, 371, 404
\bibitem[2000]{Kroupa00} Kroupa, P. 2000, astro-ph/0001202
\bibitem[1999]{Larsen99} Larsen, S. \& Richtler, T. 1999, A\&A, 345, 59
\bibitem[2001]{Larsen01} Larsen, S.S., Brodie, J., Elmegreen, B., 
   et al. 2001, ApJ, 556, 801
\bibitem[2002]{Larsen02} Larsen, S.S., Efremov, Y.N., Elmegreen, B., 
  et al. 2002, ApJ, accepted
\bibitem[1999]{Starburst99} Leitherer, C., Schaerer, D., Goldader, J.D.,
    et al. 1999, ApJS, 123, 3
\bibitem[1995]{Marlowe95} Marlowe, A.T., Heckman, T.M., Wyse, R.F.G.,
   \& Schommer, R. 1995, ApJ, 438, 563
\bibitem[1996]{Martin96}Martin, C. 1996, ApJ, 465, 680
\bibitem[1998]{Martin98} Martin, C. 1998, ApJ, 506, 222
\bibitem[2000]{Mathis00} Mathis, J.S. 2000, ApJ, 544, 347
\bibitem[2000]{Matsushita} Matsushita, S., Kawabe, R., Matsumoto, H., et al.
  2000, ApJ, 545, L107
\bibitem[1980]{Meaburn80} Meaburn, J. 1980, MNRAS, 192, 365
\bibitem[1989]{Osterbrock89} Osterbrock, D. 1989, Astrophysics of
   Gaseous Nebulae and Active Galactic Nuclei. Mill Valley:
   University Science Books.
\bibitem[1996]{Osterbrock96} Osterbrock, D.E., Fulbright, J.P., Martel, A.R.,
   et al. 1996, PASP, 108, 277
\bibitem[1998]{Paczynski98} Paczynski, B. 1998, ApJ, 494, L45
\bibitem[1999]{PZ99} Portegies Zwart, S. F., Makino, J., McMillan, S. L. W.,
   \& Hut, P.   1999, A\&A, 348, 117
\bibitem[1999]{Santi99} Santill\'an, A., Franco, J., Martos, M., \& Kim, J.
    1999, ApJ, 515, 657
\bibitem[1998]{Schlegel98} Schlegel D.J., Finkbeiner D.P., \& Douglas M.
    1998, ApJ, 500, 525
\bibitem[1999]{Selman99} Selman, F., Melnick, J., Bosch, G., \& Terlevich, R.
   1999, A\&A, 347, 532
\bibitem[1997]{Sharina97} Sharina, M.,  Karachentsev, I.D., \& Tikhonov,
    N.A. 1997, Astronomy Lett., 23, 373
\bibitem[1988]{TTB88} Tenorio-Tagle, G. \& Bodenheimer, P. 1988, ARAA, 26,
      145
\bibitem[2000]{Turatto} Turatto, M., Suzuki, T., Mazzali, P. A.
    et al. 2000, ApJ, 534, L57
\bibitem[1998]{vanZee98} van Zee, L., Salzer, J.J., Haynes, M.P.,
    O'Donoghue, A.A., \& Balonek, T.J.  1998, AJ, 116, 2805
\bibitem[1994]{Vogt94} Vogt, S.S., et al. 1994, Proc. SPIE, 2198, 362
\bibitem[1977]{Weaver77} Weaver, R., McCray, R., Castor, J., Shapiro, P.,
   \& Moore, R. 1977,  ApJ, 218, 377
\bibitem[1966]{Westerlund66} Westerlund, B.E. \& Mathewson, D.S.  1966,
    MNRAS, 131, 371
\bibitem[2000]{Zasov2000}  Zasov, A., Kniazev, A., Pustilnik, S., et al.
    2000, A\&AS, 144, 429
\end{thebibliography}
\end{document}